\documentclass{ws-ijmpd}

\begin{document}

\markboth{P. Fronczak, J. A. Ho\l yst and A. Fronczak}
{Ferromagnetic fluid as a model of social impact}

%
\catchline{}{}{}{}{}
%

\title{FERROMAGNETIC FLUID AS A MODEL OF SOCIAL IMPACT}

\author{PIOTR FRONCZAK, AGATA FRONCZAK AND JANUSZ A. HO\L YST}

\address{Faculty of Physics and Center of Excellence for
Complex Systems Research, Warsaw University of Technology,
Koszykowa 75, PL-00-662 Warsaw, Poland
\\fronczak@if.pw.edu.pl,agatka@if.pw.edu.pl, jholyst@if.pw.edu.pl}

\maketitle

\begin{history}
\received{Day Month Year}
\revised{Day Month Year}
\comby{Managing Editor}
\end{history}

\begin{abstract}
The paper proposes a new model of spin dynamics which can be treated as a model of sociological coupling between individuals. Our approach takes into account two different human features: gregariousness and individuality. We will show how they affect a psychological distance between individuals and how the distance changes the opinion formation in a social group. Apart from its sociological aplications the model displays the variety of other interesting phenomena like self-organizing ferromagnetic state or a second order phase transition and can be studied from different points of view, e.g. as a model of ferromagnetic fluid, complex evolving network or multiplicative random process. 
\end{abstract}

\keywords{Ising model; Social impact; Ferromagnetic fluids; Multiplicative random processes.}

\section{Introduction}	

Interdisciplinary research has been drawing much attention in the last decades. Models and methods developed in theoretical physics proved to be fruitful in studying complex systems \cite{haken,complex}, composed of relatively simple mutually interacting elements and coming from domains as diverge as neural networks \cite{nn}, disease spreading \cite{disease}, population dynamics \cite{popdyn}, etc. But the range of the investigations goes also beyond the natural sciences and includes problems from sociology or economy, like pedestrian motion and traffic \cite{trafic}, migrations \cite{weid,fort99} or financial crashes \cite{crashes}. 
Another important subject of this kind is the process of opinion formation in social
groups. One way of its quantitative description consists in a macroscopic approach based on the master equation
or the Boltzmann-like equations for global variables \cite{weid,weid2,helb}. Alternatively, by
making some sociologically motivated assumptions on the mechanisms of interactions
between individuals "microscopic" models are constructed and investigated numerically
or analytically by means of methods known from statistical physics \cite{galam,bahr}. One
concludes that the variety of the emerging \textit{physical} collective phenomena has much in common with the complex \textit{social} processes.

In particular, Nowak, Szamrej and Latane created a simple model based upon the successful theory of social impact in human societes first introduced by Latane in 1981 \cite{lata}. In the simplest form their model characterizes the strength of the psychological coupling between the individuals by two qualities: persuasiveness and support. The former describes the ability of one individual to persuade the other one to change his/her opinion. The later describes the ability of one individual to support the other one in his/her opinion. Different variants of the model were
explored numerically \cite{val,impact,nsl}, and many of the observations were than explained in the framework of a mean field approach \cite{lew,jh2,jh1} and the Landau theory \cite{plew}.

Here we would like to present a rather different approach to describe psychological coupling. Instead of persuasiveness and support we will study the effect of gregariousness and individuality. We will show how the two features may affect psychological distance between individuals and how the distance changes opinion formation in the society. Finally, we will show that our model could be mapped to ferromagnetic fluid not in Euclidean but in a social space.

\section{The model}

Our system consists of $N$ individuals (members of a social group); we assume that
each of them can share one of two opposite opinions on a certain subject, denoted as
$\sigma_{i}=\pm 1$, $i = 1, 2,...N$. The Hamiltonian of the model reads:

\begin{equation}
H=-\sum_{i<j} J_{i,j}(t)\sigma_{i}\sigma_{j}. 
\label{hamiltonian}
\end{equation} 

Individuals can influence each other with the strength $J_{i,j}(t)$ which can be understood as an inverse of their distance in a social space. The above means that a stronger impact corresponds to a shorter distance. We assume that social distances are changing in time and we put on the following dynamics of the strength $J_{i,j}(t)$:

\begin{equation}
J_{i,j}(t+1)=J_{i,j}(t)(1+\eta -\alpha \sigma_{i}\sigma_{j}). 
\label{coupling}
\end{equation} 

The parameter $\eta>0$ is responsible for continous growth of the social strength and can be identified as  gregariousness of $i$-th individual which leads to tightening of ties with other people. In other words, people from their nature seek the company of others. The parameter $\alpha>0$ describes another natural tendency of people which is a need to be different than a surrounding crowd, i.e. it reflects the inclination of an individual to demonstrate his/her individuality. 

For completeness of the model we assume as an initial condition any positive values of $J_{i,j}(t=0)$. The condition assures that during the system evolution couplings are always positive in the most interesting range of parameters $\eta$ and $\alpha$. 

\begin{figure}
\centerline{\psfig{file=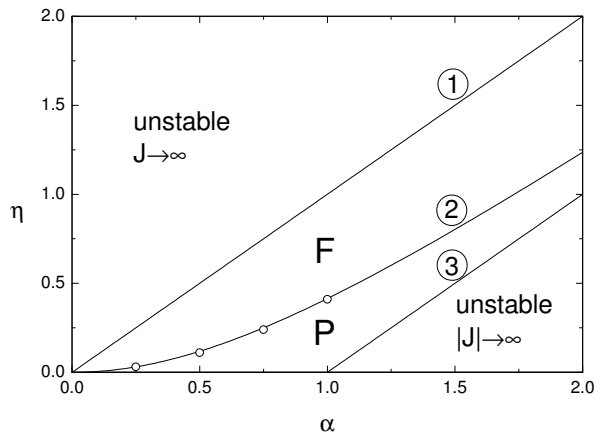,width=7.8cm}}
\caption{Phase diagram of the model (\ref{hamiltonian})-(\ref{coupling}). $F$ - ferromagnetic phase, $P$ - paramagnetic phase. Detailed explanation in the text.} \label{fig_4}
\end{figure} 

Now, let us concentrate on the phase diagram for the presented model (fig. \ref{fig_4}). It is divided into four different regions by three curves. The curve $1$ is the most obvious one. For every set of parameters above this curve, i.e. for $\eta>\alpha$, coupling strengthes will increase to infinity in exponential way. Parameters $\eta<\alpha-1$, limited by the curve $3$, also make the system unstable, but now coupling strengthes can become negative. It means that in every step $J$ will change its sign and $|J|$ will diverge to infinity. The stable region lies between two curves $1$ and $3$. 

To explain the curve $2$ let us concentrate on a single coupling and investigate the following process: $J(t)\rightarrow J(t+1)\rightarrow J(t+2)=J(t)$. As one can see from eq. \ref{coupling}, in every step $J$ grows or decreases by some well defined value. The above process is the simplest one in which $J$ stays at some fixed level, i.e. it grows and then it decreases to the same value. Let us assume that $J(t+1)>J(t)$ (the opposite case is analogous). It means that $J(t+1)=J(t)(1+\eta+\alpha)$. Then, in next step $J$ should decrease ,i.e. $J(t+2)=J(t+1)(1+\eta-\alpha)$. From above one can obtain the following condition:

\begin{equation}
\eta_{c}=\sqrt{1+\alpha^{2}}-1. 
\label{curve_2}
\end{equation} 
It is easy to see that a corresponding critical condition for a generalized process $J(t)\rightarrow J(t+1)\rightarrow J(t+2)\rightarrow ...\rightarrow J(t+2n)=J(t)$ is equivalent to the condition (\ref{curve_2}).

If we set $\eta$ below the critical value $\eta_{c}$ then the above process will result in decreasing $J$ to zero. Then it is obvious that two spins for which this coupling $J$ is investigated become disconnected and independent, what leads to paramagentic state. 

To complete discussion of the phase diagram one has to note that it is useful to assure $\alpha\ll1$. If one rewrites the eq. (\ref{coupling}) in the following form:

\begin{equation}
\frac{J_{i,j}(t+1)-J_{i,j}(t)}{\alpha J_{i,j}(t)}=\frac{\Delta J_{i,j}(t)}{J_{i,j}(t)\alpha \Delta t}=\frac{\eta}{\alpha}-\sigma_{i}\sigma_{j}, 
\label{diff}
\end{equation}  
then one can see that $\alpha$ plays the role of time scale. It means that for large $\alpha$ the succesive values of $J$ are very distant ($J$ changes very fast) and spin dynamics can not follow to compensate changes of $J$. It manifests itself in long time observed paramagnetic states interrupted by long time observed ferromagnetic states.

In summary, the interesting from the sociological point of view range of parameters is $\eta_{critical}<\eta<\alpha\ll 1$. 

The dynamics of changes of individual's opinion is given by a simple Monte Carlo procedure based on the Metropolis algorithm. A temperature $T$ given in the algorithm may be interpreted as a "social temperature" describing degree of randomness in the behavior of individuals, but also their average volatility. The procedure consists of two steps. In the first step we update states of $N$ randomly chosen individuals. In the second step we update coupling strengths for all nodes according to eq. (\ref{coupling}). 

As we will show for a wide range of parameters $\eta$ and $\alpha$, regardless of choosing a temperature the system tends to be in a ferromagnetic regime. It means that despite a tendency to manifest individuality most of individuals interact with the other people who share the same opinion.

\section{Results} \label{results}

\begin{figure} 
\centerline{\psfig{file=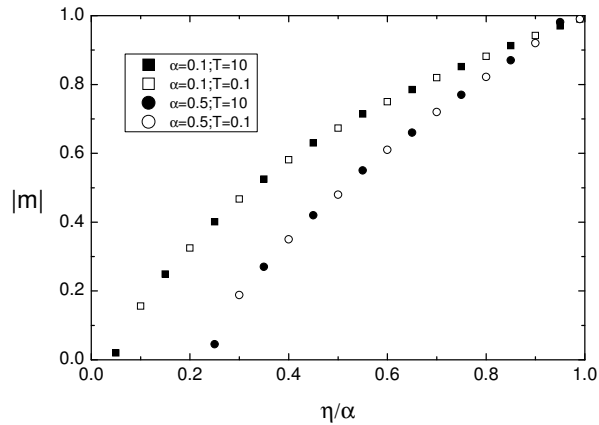,width=7.8cm}}
\caption{Dependence of average magnetization per spin on the system parameter $\eta/\alpha$.} \label{fig_1}
\end{figure}

A typical dependence of magnetization per spin $|m|$ on system parameters $\eta/\alpha$ is shown at fig. \ref{fig_1}.
Considering $\eta$ as an order parameter, continous (second order) phase transition occurs for $\eta_{c}$ given by eq. (\ref{curve_2}). Open points presented at fig. \ref{fig_4} obtained from simulations confirm that the above derivation is correct.

\begin{figure} 
\centerline{\psfig{file=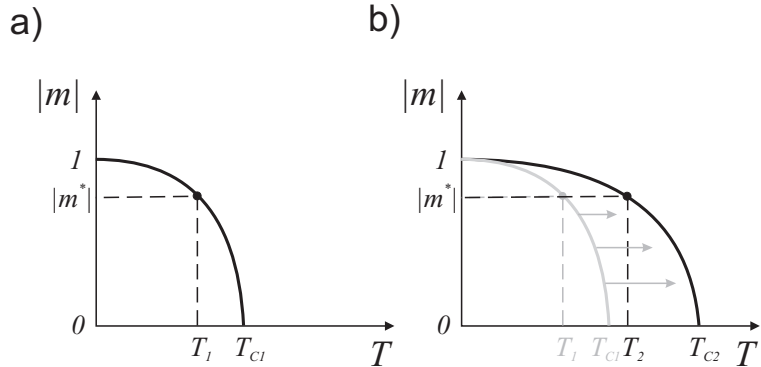,width=10cm}}
\caption{Schematic picture of processes occuring in the system during the changing of the temperature $T$ and $\eta/\alpha=const$; a) equilibrium (initial) state of the system; b) The change of the temperature $T$ from $T_{1}$ to $T_{2}$ forces magnetization curve to reshape to new equilibrium conditions.} \label{fig_2}
\end{figure}

One can see from the fig. \ref{fig_2} that the absolute value of the mean magnetization is an (increasing) function of $\eta/\alpha$ but it is completely independent on the system temperature $T$. The fact can be understood as follows. According to eq. (\ref{diff}) we can write the following equation for the mean value for the logarithm of $J_{i,j}$
\begin{equation}
\frac{1}{\alpha}\left\langle\frac{d\ln J_{i,j}}{ dt}\right\rangle=\frac{\eta}{\alpha}-\langle\sigma_i\sigma_j\rangle.
\label{jh1}
\end{equation}

However, the mean value of the product $\langle\sigma_i\sigma_j\rangle$ is related to the mean system magnetization $\langle m\rangle$ which on the other hand is a certain function of the Boltzmann factors $\exp(-J_{i,j}/T)$. Thus taking andvantage of the mean field approximation we can write that 
\begin{equation}
\frac{1}{\alpha}\left\langle\frac{d\ln J_{i,j}}{dt}\right\rangle=\frac{\eta}{\alpha}-g(\langle J_{i,j}\rangle/T).
\label{jh2}
\end{equation}
When the system is close to equilibrium the left hand side of the last equation in average equals to zero and the equation simplifies to the following relation 
\begin{equation}
\langle J_{i,j}\rangle/T = g^{-1}(\eta/\alpha).
\label{jh3}
\end{equation}
In this sense the average value of the coupling constant is always proportional to the temperature and a function of a ratio $\eta/\alpha$.  

\begin{figure}
\centerline{\psfig{file=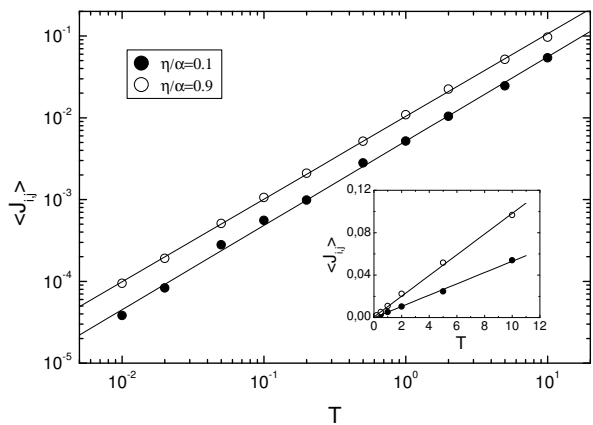,width=7.8cm}}
\caption{Average coupling strength in the system as a function of temperature. Log-log fit is presented by straight lines with slopes $1.03$ for open points and $1.01$ for filled points. Inset: the same dependence in linear scale. Slopes of two fitted linear functions represent a proportionality factor given by function $g^{-1}(\eta/\alpha)$ from eq. (\ref{jh3}).} \label{fig_new}
\end{figure}

Since $m$ is a function of the ratio $\langle J_{i,j}\rangle/T$ thus it only depends on the ratio $\eta/\alpha$ and does not depend on the system temperature. The numerical confirmation of the statement is presented at fig. \ref{fig_new}. 

To illustrate the effect one can perform the following procedure. First, set values of the parameters $\eta$ and $\alpha$, and also temperature $T_{1}$ of the system. (These values result in a given spin magnetization depicted in fig. \ref{fig_2} by $m^{*}$, and they also correspond to a certain distribution of couplings between spins $P(J)$.) Now we would like to reconstruct the whole magnetization curve passing through the point $(T_{1},m^{*})$. Unfortunately, as we have already shown any change of $T$ modifies coupling strengthes, which in consequence modify the shape of the magnetization curve. Reconstruction of $m(T)$ will be possible if we \textit{freeze} $P(J)$, i.e. we make each $J_{i,j}$ constant. 

Once we have determined $m(T)$, we restore dependence of $J$ on $T$ and change temperature to new value $T_{2}>T_{C1}$. A new pair $(T_{2},m^{*})$ determines a new critical temperature $T_{C2}$. The curve at the fig. \ref{fig_2}a adjusts to new conditions and transforms to the shape shown at the fig. \ref{fig_2}b (obtained by the same method as before). It means that regardless of a choice of the system temperature we are always below the critical temperature, i.e. in the ferromagnetic state.

The sociological conclusion could be as follows: regardless of a "social temperature" people always try to correlate their opinions with others (create groups of interest). This tendency to share the same opinion with other people, regardless of some external forces, make us, people, so resistant to trials of despots to make the people unorganized and disoriented. Of course the parameter $\eta/\alpha$ characterizes our own (not social) point of view which gives us some independency respecting other people opinion.

\begin{figure} 
\centerline{\psfig{file=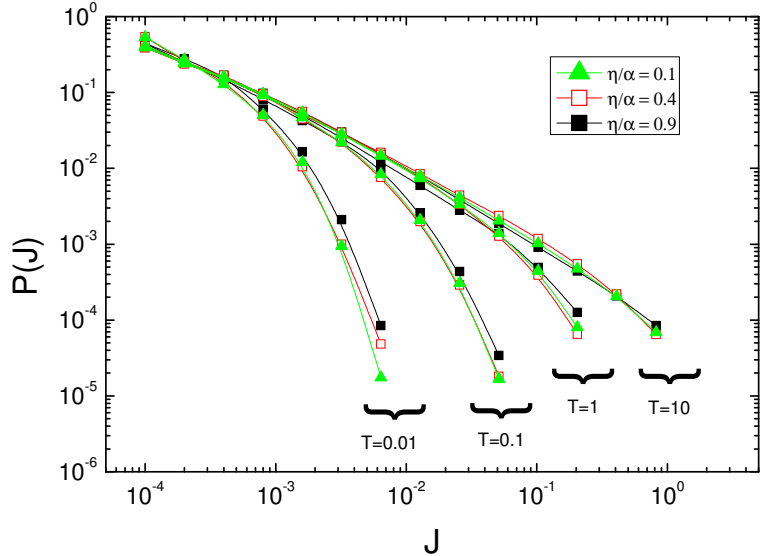,width=10cm}}
\caption{Distribution of coupling strengths for different values of parameters $\eta/\alpha$ and $T$.} \label{fig_3}
\end{figure}

From the point of view of complex networks domain \cite{dorogovtsev,newman} it is interesting to consider the model as a weighted network, where nodes correspond to individuals and links have assigned weights equal to coupling strength. One of nontrivial observations is a distribution of coupling strengthes $P(J)$ which is presented at fig. \ref{fig_3}. As one can see for large temperature $T$ the distribution has a form of power law with the exponent $\gamma\approx 0.85$. 

It seems that there should be a strong relation between the observed power-law distributions and distributions obtained due to a more general class of multiplicative random processes \cite{levy,sornette}. If fact, one can easily find some similarity of eq. (\ref{coupling}) to eq. (1) in \cite{levy}. The differences occur when one takes into account the temperature and its influence on distributions at fig. (\ref{fig_3}). We suspect that the model studied by us settles somewhere between two multiplicative random processes studied in \cite{levy} and \cite{sornette}. This hypothesis is still under investigation and the results will be published elsewhere.

Now let us draw attention to similarity of the presented model to magnetic fluids which are widely studied for the last thirty years \cite{fluids}. Magnetic fluids are described by interacting molecules with both translational and spin degrees of freedom. They interact due to weak long-ranged exchange interactions in addition to spin-independent isotropic attractive forces. The most simple physical parameter used in phase diagrams of magnetic fluids has a form

\begin{equation}
R = \frac{\int\phi_{ex}(\vec{r})d\vec{r}}{\int\phi_{attr}(\vec{r})d\vec{r}}, 
\label{fluid}
\end{equation} 

where $\phi_{ex}(\vec{r})$ describes exchange integral, $\phi_{attr}(\vec{r})$ some attractive-type integral and $\vec{r}$ is Euclidean distance between molecules. One can easily see a correspondence between the above parameter and the main parameter of our model $\eta/\alpha$. In that sense $\eta$ and $\alpha$ have respectively attractive and spin-dependent properties, and our distance in social space reflects Euclidean distance $\vec{r}$. 

\section{Conclusions} \label{conclusions}

In this paper we propose a new model of spin dynamics which could be treated as a model of sociological coupling between individuals. Apart from its sociological aplications the model displays the variety of other interesting phenomena like self-organizing ferromagnetic state or a second order phase transition and can be studied from different points of view, for example as a model of ferromagnetic fluid, complex evolving network or multiplicative random process. 

\section*{Acknowledgments}

The work has been supported by European Commission Project CREEN
FP6-2003-NEST-Path-012864. A.F. acknowledges financial support from the Foundation for Polish Science (FNP 2006).


\begin{thebibliography}{00}  

\bibitem{haken}  H. Haken, {\it Synergetics. An Introduction} (Springer-Verlag, Heidelberg, New~York, 1983);  {\it Advanced  Synergetics} (Springer-Verlag, Heidelberg, New~York, 1983). 
\bibitem{complex} G. A. Cowan, D. Pines, D. Meltzer (eds.), {\it Complexity. Metaphors, Models, and Reality} (Addison-Wesley, Santa Fe, 1994).
\bibitem{nn} D. Amit, {\it Modeling Brain Function} (Cambridge Univ. Press, Cambridge, 1989);E. Domany, J.L. van Hemmen, K. Schulten (eds.) {\it Models of Neural Networks} (Springer, Berlin, 1995); A. Browne (ed.), {\it Neural network analysis, architectures and applications} (Institute of Physics Publishing, Bristol, 1997).
\bibitem{disease} A. Johansen, {\it Physica D} {\bf 78},186 (1994); H. C. Tuckwell, L. Toubiana, J-F. Vibert, {\it Phys. Rev. E} {\bf 57}, 2163 (1998).
\bibitem{popdyn} P. Bak, K. Sneppen, {\it Phys. Rev. Lett.} {\bf 71}, 4083 (1993); A. Pekalski, {\it Physica A} {\bf 252}, 325 (1998);
\bibitem{trafic} D. Helbing, {\it Phys. Rev. E} {\bf 55}, 3735 (1997); {\it Physica A} {\bf 219}, 375 (1995); D. Helbing, P. Molnar, {\it Phys. Rev. E} {\bf 51}, 4282 (1995). 
\bibitem{weid} W. Weidlich, G. Haag, {\it Concepts and Models of Quantitatively Sociology} (  Springer, Berlin, New~York, 1983); W. Weidlich, {\it Physics Reports} {\bf 204}, 1 (1991).
\bibitem{fort99} J. Fort, V. M\'{e}ndez, {\it Phys. Rev. Lett.} {\bf 82}, 867 (1999).
\bibitem{crashes} D. Sornette, A. Johansen, {\it Physica A} {\bf 245}, 1 (1997); N. Vandewalle, M. Ausloos, P. Boveroux, A. Minguet, {\it Eur. Phys. J. B} {\bf 4}, 139 (1998).
\bibitem{weid2} W. Weidlich, J. Math. {\it Sociology} {\bf 18}, 267 (1994).
\bibitem{helb} D. Helbing, {\it Physica A} {\bf 193}, 241 (1993); {\it J. Math. Sociology} {\bf 19}, 189 (1994); D. Helbing, {\it Quantitative Sociodynamics} (Kluwer Academic, Dordrecht, 1995). 
\bibitem{galam} S. Galam, {\it Physica A} {\bf 230}, 174 (1996); {\it Physica A} {\bf 238}, 66 (1997).
\bibitem{bahr} D. B. Bahr, E. Passerini, {\it J. Math. Sociology} {\bf 23}, 1 (1998).
\bibitem{lata}  B. Latan\'{e},  {\it Am. Psychologist} {\bf 36}, 343 (1981). 
\bibitem{val}  R. R. Vallacher, A. Nowak (Eds.), {\it Dynamical systems in social psychology} ( San Diego, Academic Press, 1994). 
\bibitem{impact} E. L. Fink, {\it J. Communication} {\bf 46}, 4 (1996); B. Latan\'{e}, {\it J. Communication} {\bf 46}, 13 (1996). 
\bibitem{nsl}  A. Nowak, J. Szamrej, B. Latan\'{e}, {\it Psych. Rev.} {\bf 97}, 362 (1990). 
\bibitem{lew}  M. Lewenstein, A. Nowak, B. Latan\'{e }, {\it Phys. Rev. A} {\bf 45}, 763 (1992).
\bibitem{jh2} K. Kacperski and J.A. Holyst, {\it J. Stat. Phys.} {\bf 84}, 169 (1996).  
\bibitem{jh1} J.A. Holyst, K. Kacperski and F. Schweitzer, {\it Annual Review of Comput. Phys.} {\bf 9}, 253-273 (2001).
\bibitem{plew} D. Plewczy\'nski, {\it Physica A} {\bf 261}, 608 (1998).
\bibitem{dorogovtsev} S.N. Dorogovtsev and J.F.F. Mendes, {\it Evolution of networks} (Oxford Univ.Press, 2003). 
\bibitem{newman} S. Bornholdt and H.G. Schuster, {\it Handbook of graphs and networks} (Wiley-Vch 2002). 
\bibitem{levy} M. Levy and S. Solomon, {\it Int. J. Mod. Phys. C} {\bf 7}, 595 (1996).
\bibitem{sornette} D. Sornette and R. Cont, {\it J. Phys. I (France)} {\bf 7}, 431 (1997).
\bibitem{fluids} P. C. Hemmer and D. Imbro, {\it Phys. Rev. A} {\bf 16}, 380 (1977); J. M. Tavares et al, {\it Phys. Rev. E} {\bf 52}, 1915 (1995); W. Fenz and R. Folk, {\it Phys. Rev. E} {\bf 67}, 021507 (2003); F. Lado and E. Lomba, {\it Phys. Rev. Lett.} {\bf 80}, 3535 (1998).
\end{thebibliography}
\end{document}